\documentclass[letterpaper,11pt]{AAS}	

\usepackage{bm}
\usepackage{amsmath}
\usepackage{subfigure}
\usepackage[colorlinks=true, pdfstartview=FitV, linkcolor=black, citecolor= black, urlcolor= black]{hyperref}
\usepackage{overcite}
\usepackage{footnpag}			      	

\PaperNumber{XX-XXX}

\begin{document}

\title{Rendezvous Mission for Interstellar Objects Using a Solar Sail-Based Statite Concept}

\author{Richard Linares\thanks{Charles Stark Draper Assistant Professor, Department of Aeronautics and Astronautics, Massachusetts Institute of Technology, Cambridge, MA 02139, Email: linaresr@mit.edu}, Damon Landau\thanks{Mission Formulation Engineer, Project Systems Engineering \& Formulation Section, Jet Propulsion Laboratory, California Institute of Technology, Pasadena, CA 91109, Email: { damon.landau@jpl.nasa.gov}}, Daniel Miller \thanks{Ph.D. Student, Massachusetts Institute of Technology, Cambridge, MA 02139, Email: dmmiller@mit.edu}, Benjamin Weiss \thanks{Professor, Department of Earth, Atmospheric
and Planetary Sciences, Massachusetts Institute of Technology, Cambridge, MA 02139.}, and Paulo Lozano  \thanks{Professor, Department of Aeronautics and Astronautics, Massachusetts Institute of Technology, Cambridge, MA 02139.}
}

\maketitle{}

\begin{abstract}
Using the ``statite," or static-satelite, concept -- an artificial satellite capable of hovering in place using a solar sail -- this work proposes to create a dynamic orbital slingshot in anticipation of Interstellar Objects (ISOs) passing through our solar system. The existence of these ISOs offers a unique scientific opportunity to answer fundamental scientific questions about the origin of solar system volatiles, the compositions of exo-solar systems, and the transfer rates of material between solar systems. However, due to their high heliocentric velocities and relatively short lead time, it may be extremely difficult to visit ISOs with current satellite propulsion systems. This work investigates the statite concept as applied to ISO missions and demonstrates potential configurations for optimal ISO flyby and rendezvous missions.
\end{abstract}

\section{INTRODUCTION}

Visiting Interstellar Objects (ISOs) is a grand engineering challenge. The study of asteroids and comets has revealed a treasure trove of information about the formation and history of our solar system. In 2017, the first ISO \textit{‘Oumuamua} was discovered \cite{ref1} and recently, in August 2019, a second object was discovered \cite{ref2} which flew by the Sun in December 2019. The existence of these ISOs offers us a great scientific opportunity to develop a detailed understanding of the formation and history of other star systems. In this work, a Dynamic Orbital Slingshot concept is proposed for a rapid flyby and rendezvous mission to an ISO (e.g., Interstellar asteroids, comets). Studying ISOs up close will be critical to answering fundamental scientific questions. However, due to their high heliocentric velocities and relatively short lead time, this may be extremely difficult with current satellite propulsion systems. This work proposes the use of a statite \cite{ref3}, an artificial satellite that employs a solar sail to continuously hover in place without entering into an orbit. In essence, the solar sail is stationary in the heliocentric frame and, once an ISO is detected, the system can deliver a CubeSat on either a flyby or rendezvous trajectory (with or without propulsion). 

\begin{figure}
	\centering
	\includegraphics[width=0.8\textwidth]{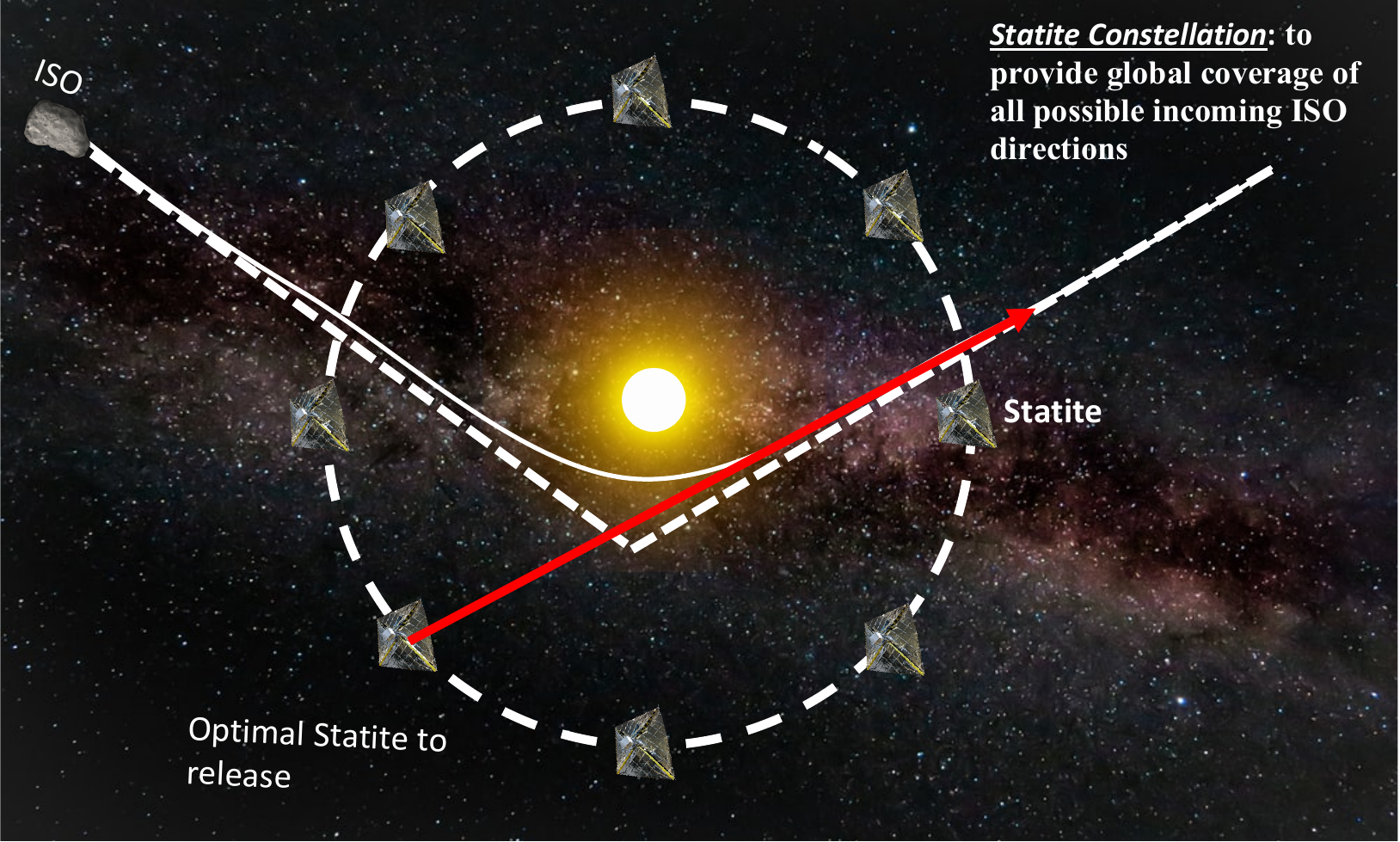}
	\caption{The mission concept: Using a constellation of Statites for global coverage (although a mission using a single statite is possible)}
	\label{mission concept}
\end{figure}

The proposed mission concept is to have a constellation of statites where: A) each statite enters into a stationary, “low-energy” state, which it can hold indefinitely,  awaiting a potential ISO; B) once an ISO is detected, a flyby or rendezvous trajectory is calculated with an expected 4-16 months of lead time; C) a single statite releases a CubeSat which enters into a freefall trajectory with respect to the Sun or uses attitude control to orient the solar sail and adjust the solar radiation pressure force to accomplish a rendezvous; D) the CubeSat, which may use propulsion or the solar sail, then adjusts its trajectory for a flyby or rendezvous with the ISO; and E)  onboard sensors are used to make critical scientific measurements. In phase I, we will investigate the feasibility of the proposed approach; in particular, we will: 1) investigate the statite constellation design and determine potential configurations for optimal ISO flybys; 2) study the ability of the system to achieve a rapid rendezvous; 3) investigate the addition of solar electric propulsion for improving rendezvous performance; 4) determine size requirements for potential instruments that will be required onboard for making scientific measurements; and 5) finally, develop a detailed concept of operations for a mission that has the highest possible scientific benefit.



\subsection{Previous Work} 
The use of solar sail technologies has been extensively studied \cite{ref3,ref4,ref5,ref6,ref7,ref8,ref9,ref10,ref11}. Many studies have investigated orbital dynamics of solar sail spacecraft and they have found several types of stable orbits, including terminator orbits \cite{ref9}, quasi-terminator orbits, heliotropic orbits \cite{ref4,ref5}, and Delta-V assisted periodic orbits \cite{ref10}.

\section{Trajectory Calculations}

\begin{figure}[b]
	\centering
	\includegraphics[width=0.4\textwidth]{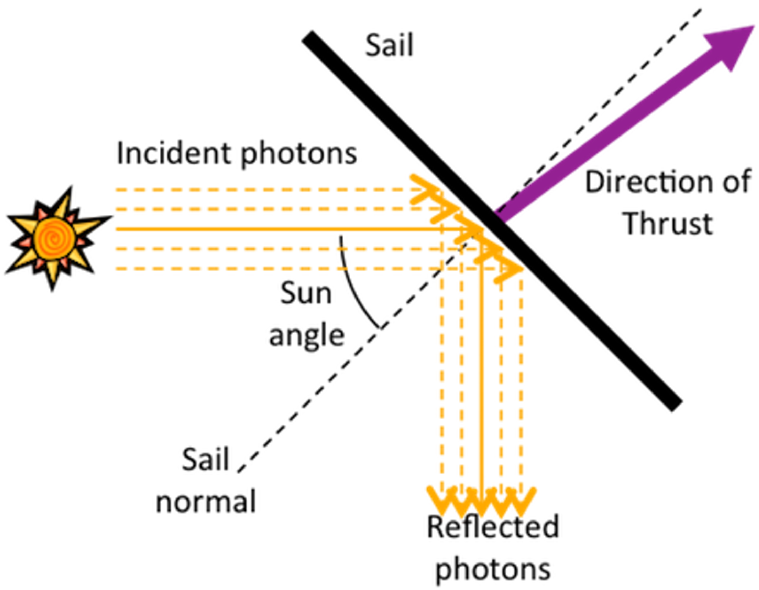}
	\caption{Solar Sail Reflection Geometry 
(source: www.deorbitsail.com)}
	\label{fig:geometry}
\end{figure}

The proposed concept uses solar sail technology in a novel way to allow for rapid and responsive flyby and rendezvous missions. Central to the innovation of the proposed work is the concept of a statite, or static satellite, first proposed by Forward \cite{ref7}. Such a spacecraft would use a solar sail to “cancel” out the gravitational acceleration caused by the Sun \cite{ref3,ref4,ref5,ref6,ref7,ref8} by reflecting light to generate a propulsive force. This process is a function of the area-to-mass ratio of the sail and the distance to the Sun. This latter detail is critical, for a key idea behind the concept of a statite is that both the gravitational force of the Sun and the propulsive force from the solar sail vary as $1/R^2$. This allows the statite to exactly cancel the gravitational force hover at any distance from the sun, but it must have an area-to-mass ratio greater than a critical value. It turns out that the ability of the statite to “cancel” out the gravitational acceleration of the Sun is independent of the distance of the sail to the Sun. The key parameter to achieve a statite is the area-to-mass ratio, and the critical value is determined by the balance between gravitational attraction and light pressure repulsion using the following \cite{ref7}: 
\begin{equation}
F=\frac{GM_\text{sun}m}{R^2}=2\frac{AL}{4\pi R^2C}
\end{equation}
where $M_\text{sun}$ is the mass of the Sun, $A$ is the area of the sail, and $L$ is the total light flux from the Sun at $R=1$ (in units of AU=$1.5\time 10^{11}$ m), which is $L \approx1400$ W/m$^2$. The reflection geometry is shown in Figure \ref{fig:geometry}. The mass-to-area ratio is then given by
\begin{equation}
\frac{m}{A}=\frac{L}{2\pi C GM_\text{sun}m}
\end{equation}

Using this equation, the critical mass-to-area ratio\cite{ref3,ref4,ref5,ref6,ref7} is 1.6 g/m$^2$  and the area-to-mass ratio is 0.625 m$^2$/g. To achieve statite-level performance, the payload and solar sail combination of mass $m$ must provide these meet these specifications. This may be achievable with modern materials using “aluminized, temperature-resistant material,” also called CP-1\cite{ref6}. 

Since the solar sail is levitating or hovering, its inertial velocity is zero. When it is released from this state, it will enter into a Keplerian orbit with the following period:
\begin{equation}
E=-\frac{\mu}{2}+\frac{1}{2}v^2=-\frac{2\mu}{a}
\end{equation}
In this context, $\mu = \mu_\text{sun}=GM_\text{sun}=4\pi^2$ in units of AU and years. If we assume $a=r/2$, we get the following relationship for the period of the trajectory:
\begin{equation}\label{period}
T=2\pi\sqrt{\frac{r^3}{8\mu}}
\end{equation}
Assuming that the statite is released at 1 AU, the time until a hypothetical collision with the Sun is 0.1768 years. 
Note in Eq.~\ref{period} half of the period is used since the payload is in free-fall once released and only “flies” half the trajectory (the other half occurs after a hypothetical collision with the Sun). This relationship shows that a statite at 1 AU has a free-fall trajectory of about 64 days. This fast response time to a potential ISO can be thought of as a slingshot effect, since the solar sail is used to “store energy” that is released when desired. Additionally, to achieve a flyby some Delta-V is required to adjust from the free-fall path to a flyby trajectory. The proposed mission  for the statite concept is to utilize a constellation of such devices to achieve wider coverage over a spherical region of 1 AU for potential ISO missions. Additionally, the orbital plane can be adjusted with relatively low Delta-V. Note that because of the constant propulsive thrust capability of solar sails, the resultant force vector can be used to cancel out a portion of the solar gravity pull, making the effective net attraction adjustable (effectively adjusting $\mu_\text{sun}$), and these adjustments are independent of orbital radius. These types of orbits are termed quasi-satellite or Quasites. Exotic trajectories can thus be explored to balance the responsiveness of the dynamic slingshot with other design parameters, such as the Delta-V required. As shown in Figure 1, a number of statites could be placed in hovering orbits around the Sun in a standby state until an ISO is detected. Others could be placed over each pole or in solar synchronous orbits with a period selected based on design parameters.

\section{Simulation Results}
Flyby trajectories are an important part of this study and have very simple mission Concept of Operations (CONOPS), but, in terms of scientific value, we consider the rendezvous mission to be the “holy grail” for ISO science. To better understand the feasibility of rendezvous our team has conducted some initial proof-of-concept simulations. 
\begin{figure}[h]
	\centering
	\includegraphics[width=0.6\textwidth]{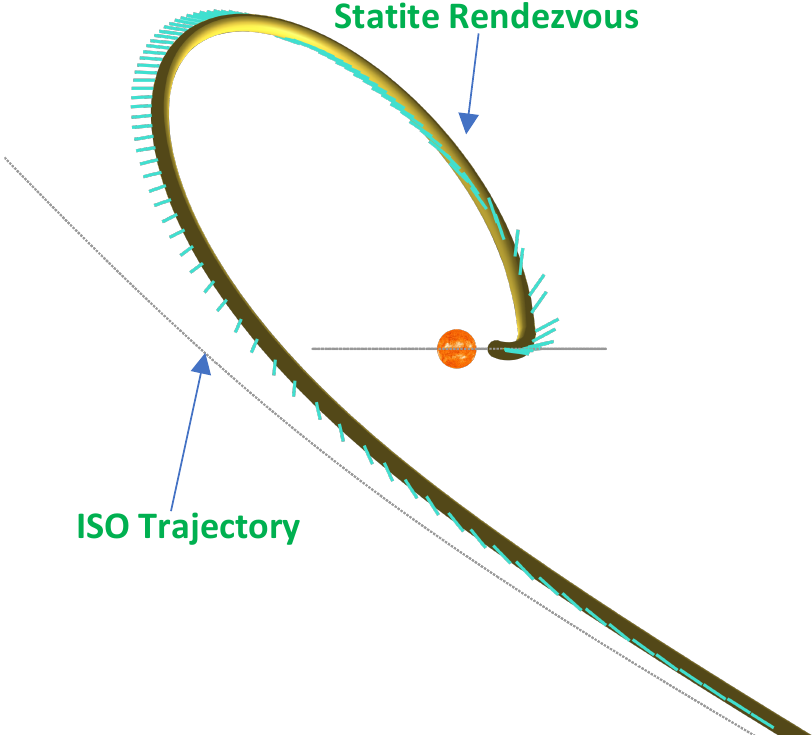}
	\caption{Rendezvous with \textit{Borisov}}
	\label{figure2a}
\end{figure}

\begin{figure}[h]
	\centering
	\includegraphics[width=0.8\textwidth]{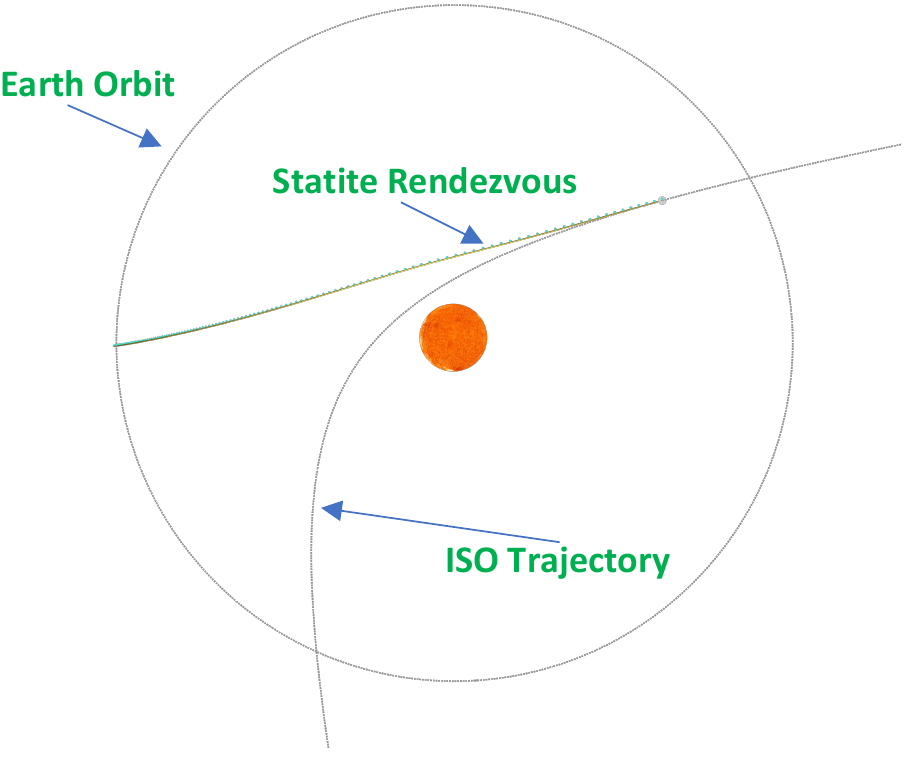}
	\caption{Rendezvous with \textit{‘Oumuamua}}
	\label{figure2b}
\end{figure}

Figures \ref{figure2a} and \ref{figure2b}  demonstrate the rendezvous concept utilizing statites for both the \textit{‘Oumuamua} and \textit{Borisov} trajectories. These simulations were produced using the ZoSo optimizer, one of Jet Propolusion Lab's (JPL’s) trajectory optimization software tools \cite{ref13}. The initial, stationary states from which the trajectories began were required to meet the same criteria: a location 1 AU from the Sun in the ecliptic. While this approximately matches the orbital path of the Earth, the statites' starting positions did not match that of the Earth; the departure time was free. Each statite was released from its stationary configuration by controlling the inertial normal direction of the solar sail. Figure \ref{figure2a} shows the normal directions (blue vectors) during rendezvous with \textit{Borisov}, while the trajectory to 'Oumuamua is shown in Figure \ref{figure2b}. These initial proof-of-concept simulations show great promise for this concept. Indeed, from our initial proof-of-concept studies, we have concluded that very high DeltaV missions are possible using statites. 

\begin{table}
\centering
\begin{tabular}{|lllll|}
\hline
Target & Objective & Depart & Rendezvous & DeltaV \\
\hline
 ISO1 (\textit{‘Oumuamua}) & Max. departure time & Jun 25, 2017 & Oct 1, 2017 & 104 km/s \\
  ISO2 (\textit{Borisov}) & Min. flight time & Dec 29, 2016 &  Feb 24, 2024 & 145 km/s\\
 \hline
\end{tabular}
	\caption{Rendezvous Trajectory Optimization Results for \textit{‘Oumuamua} and \textit{Borisov} using a solar sail with statite-level performance. The Delta-V is generated using solar radiation pressure force.}\label{table1}
\end{table}

From Table \ref{table1}, it can be seen that a DeltaV of 104 km/s and 145 km/s are achieved for the \textit{'Oumuamua} and \textit{Borisov} rendezvous trajectories, respectively. These high Delta-V values are possible due to the fact that the statite can cancel the solar gravity completely, independent of the distance to the Sun. Moreover, from these initial proof-of-concept results, we can see that the rendezvous times are relatively short for both of these notional missions. For \textit{'Oumuamua}, the rendezvous time is approximately 5 months and the rendezvous occurred at 0.76 AU. For \textit{Borisov}, the rendezvous time was approximately 8 years and the rendezvous occurred at 30.6 AU. Although the \textit{Borisov} trajectory had a long time of flight, it was still within typical mission timescales, while requiring neither large propulsion systems nor long lead times. 

The differences between these two examples are interesting and will motivate further the technical and theoretical studies. Based on these preliminary results, \textit{‘Oumuamua} would have been easier to intercept than \textit{Borisov}, given the same starting configuration of the statite. While the reasons for this are not completely understood, the authors propose the following hypothesis. From Figure \ref{figure2b}, \textit{‘Oumuamua} appears to be aligned with the "free-fall" trajectory and the line connecting the statites initial location to the Sun is close to the asymptote of the hyperbolic trajectory of the ISO. This suggests an optimal statite placement for future rendezvous trajectories. The authors would therefore suggest a small constellation of statites to allow for global coverage of all potential hyperbolic orbital asymptote directions. Additionally, as demonstrated in Figure \ref{figure2a}, statites can easily rendezvous with an ISO whose trajectory is not aligned with the ecliptic. This is due to the fact that in the stationary state, the statite has zero angular momentum and, once released, it can use the solar sail to enter into a heliocentric trajectory in any orbital plane. 

The high Delta-V achieved during these transfers is indeed an important selling point for the statite concept. As a propellant-less form of propulsion, a solar sails provides the statite with essentially “free” Delta-V. As a result, a high Delta-V vs a low Delta-V trajectory is not a critical discriminator for achieving rendezvous or flybys. Instead, the most important design metric for solar sail missions is flight time. Indeed, even in the stationary configuration, statites will accumulate high Delta-V numbers. For example, they accumulate 187 km/s per year just sitting at 1 AU (countering the force of the Sun). Catching an interstellar object is an inherently high-Delta-V operation, and statites are particularly suited to accomplish this.

\section{Other Applications}

The proposed concept may have additional applications outside of ISO missions. In the past, statites have been suggested for Earth orbiting missions \cite{ref4,ref5}. Not considered in these applications is the ability of statites to rendezvous rapidly with objects in diverse orbital planes. This capability could be put to use accomplishing on-orbit satellite servicing and space debris removal, two missions in which the location of a potential target satellite may be unknown at the start of the mission. The statite concept could also be used to design a constellation of satellite servicers. 

Finally, given the proof-of-concept rendezvous results demonstrating the ability of statites to quickly enter Sun escape trajectories, these spacecraft may be useful for deep space missions to outer planets or asteroids. Figure 5 shows a simple calculation conducted by the authors to determine the potential escape speed achievable using a statite. Consider a spacecraft that free falls towards the Sun from an initial position at 1 AU. Upon reaching perihelion, it orients its solar sail away from the Sun to maintain the speed that it gained during the fall. Using this strategy, the spacecraft could achieve speeds of up to 25 AU/year. For comparison, Voyager 1 only attained 3.6 AU/yr. The achievable escape speed is, however, a function of the perihelion distance and the proximity to the Sun necessary for achieving the highest speeds would create thermal concerns. While this  analysis is rudimentary -- it does not  account for angular momentum -- it shows what is achievable from an energy standpoint..

\pagebreak

\begin{figure}[h]
	\centering
	\includegraphics[width=0.6\textwidth]{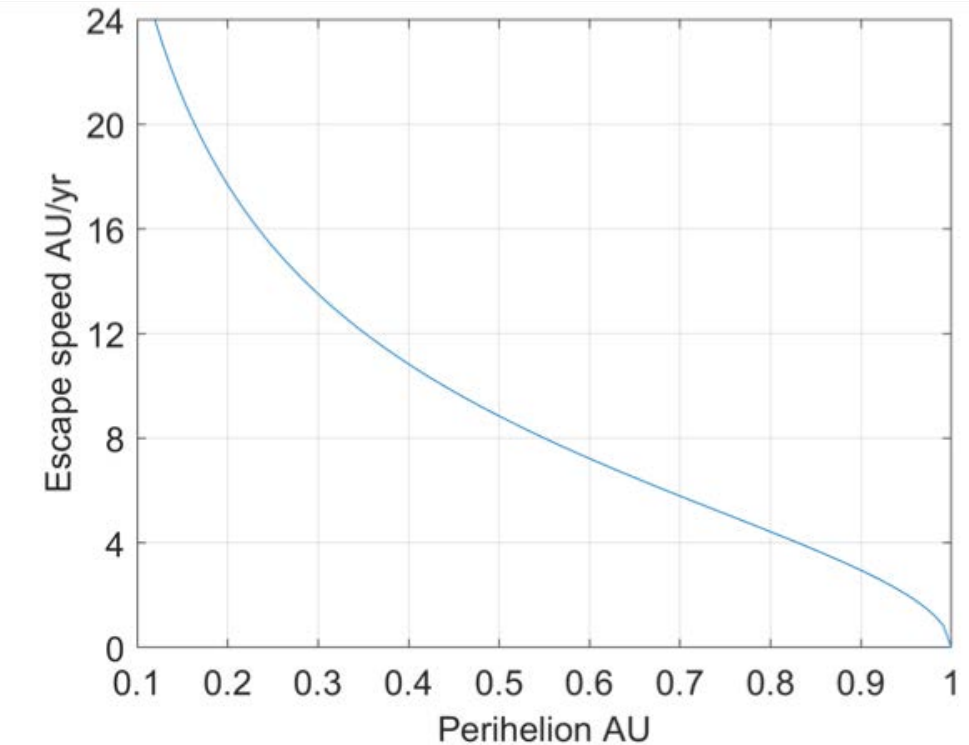}
	\caption{Solar System escape velocity for Statites.}
	\label{geometry}
\end{figure}

\section{Conclusion}

ISOs are a newly discovered class of objects with great scientific value. Unfortunately, due to their high heliocentric velocities and the difficultly of discovering them with significant lead time, they are difficult targets to study up close using spacecraft with current propulsion technology. Using the proposed solar sail statite concept, a solution to this problem may be found. This work demonstrated that these spacecraft are able to use their solar sails to hover in place while awaiting an ISO and, once a target is found, rapidly accelerate to the velocity necessary to achieve a rendezvous. This paper examined two cases corresponding to the two discovered ISOs: \textit{'Oumuamua} and \textit{Borisov}. In both scenarios, it was shown that a solar sail spacecraft with statite-level performance could achieve a rendezvous. Additionally, it was noted that the geometry of these trajectories impacts the time of flight of the mission. These initial proof-of-concept results are promising, but additional investigation is required. 



\section{Acknowledgment}
Part of the research was carried out at the Jet Propulsion Laboratory, California Institute of Technology, under a contract with the National Aeronautics and Space Administration \linebreak (80NM0018D0004). The information in this document is pre-decisional and is provided for planning and discussion only.

\bibliographystyle{AAS_publication}   
\bibliography{references}   

\begin{thebibliography}{10}

\bibitem{ref1}
E.~Mamajek, ``Kinematics of the Interstellar Vagabond 1I/'Oumuamua (A/2017
  U1),''  {\em arXiv preprint arXiv:1710.11364}, 2017.

\bibitem{ref2}
J.~O'Callaghan, {\em A Second Interstellar Object May Be Streaking through Our
  Solar System}, Sep 13, 2019 (accessed May 21, 2020).

\bibitem{ref3}
R.~L. Forward, ``Statite: spacecraft that utilizes sight pressure and method of
  use,''  Feb.~2 1993.
\newblock US Patent 5,183,225.

\bibitem{ref4}
C.~R. McInnes and J.~F. Simmons, ``Solar sail halo orbits part II-geocentric
  case,''  {\em Journal of Spacecraft and Rockets}, Vol.~29, No.~4, 1992,
  pp.~472--479.

\bibitem{ref5}
C.~R. McInnes and J.~F. Simmons, ``Solar sail halo orbits part II-geocentric
  case,''  {\em Journal of Spacecraft and Rockets}, Vol.~29, No.~4, 1992,
  pp.~472--479.

\bibitem{ref6}
S.~Baig and C.~R. McInnes, ``Light-levitated geostationary cylindrical orbits
  are feasible,''  {\em Journal of Guidance, Control, and Dynamics}, Vol.~33,
  No.~3, 2010, pp.~782--793.

\bibitem{ref7}
R.~L. Forward, ``Advanced Space Propulsion Study-Antiproton and Beamed Power
  Propulsion,''  tech. rep., HUGHES RESEARCH LABS MALIBU CA, 1987.

\bibitem{ref8}
D.~Lantukh, R.~P. Russell, and S.~Broschart, ``Heliotropic orbits at oblate
  asteroids: balancing solar radiation pressure and J2 perturbations,''  {\em
  Celestial Mechanics and Dynamical Astronomy}, Vol.~121, No.~2, 2015,
  pp.~171--190.

\bibitem{ref9}
D.~Lantukh, R.~P. Russell, and S.~B. Broschart, ``Heliotropic Orbits at
  Asteroids: Zonal Gravity Perturbations and Application at Bennu,''  {\em
  Advances in the Astronautical Sciences, Spaceflight Mechanics 2015},
  Vol.~155, 2015.

\bibitem{ref10}
S.~Kikuchi, Y.~Tsuda, and J.~Kawaguchi, ``Delta-V assisted periodic orbits
  around small bodies,''  {\em Journal of Guidance, Control, and Dynamics},
  2017, pp.~150--163.

\bibitem{ref11}
E.~Morrow, D.~J. Scheeres, and D.~Lubin, ``Solar sail orbit operations at
  asteroids,''  {\em Journal of Spacecraft and Rockets}, Vol.~38, No.~2, 2001,
  pp.~279--286.

\bibitem{ref13}
D.~Landau, ``Efficient maneuver placement for automated trajectory design,''
  {\em Journal of Guidance, Control, and Dynamics}, Vol.~41, No.~7, 2018,
  pp.~1531--1541.

\end{thebibliography}

\end{document}